\documentclass[12pt]{iopart}
\usepackage{mathrsfs, amssymb}
\usepackage{graphicx}
\usepackage{hyperref}
\usepackage[english]{babel}
\usepackage{braket}
\usepackage{booktabs}

\begin{document} \title{The  quintessence field  as a  perfect cosmic  fluid of
constant      pressure}      \author{Wenzhong      Liu$^{\heartsuit}$,      Jun
Ouyang$^{\heartsuit}$     and    Huan-Xiong     Yang$^{\heartsuit,\spadesuit}$}
\address{$^{\heartsuit}$Interdisciplinary   Center   for   Theoretical   Study,
University of  Science and Technology  of China,  Hefei, 200026, P.  R. China\\
$^{\spadesuit}$Kavli  Institute for  Theoretical  Physics  China, CAS,  Beijing
100190,    China}    \ead{wzhliu@mail.ustc.edu.cn,    yangjun@mail.ustc.edu.cn,
hyang@ustc.edu.cn} \date{\today}

\begin{abstract} We study the cosmology of a quintessence scalar field which is
equivalent  to  a  non-barotropic  perfect fluid  of  constant  pressure.   The
coincidence problem is alleviated by such a quintessence equation-of-state that
interpolates between  plateau of zero at  large redshifts and plateau  of minus
one as  the redshift approaches to  zero.  The quintessence field  is neither a
unified  dark matter  nor  a mixture  of cosmological  constant  and cold  dark
matter,  relying  on the  facts  that  the  quintessence density  contrasts  of
sub-horizon modes would  undergo a period of late-time decline  and the squared
sound speeds of quintessence perturbations do not vanish.  What a role does the
quintessence  play is  dynamic  dark energy,  its  clustering could  remarkably
reduce  the  growth  rate  of the  density  perturbations  of  non-relativistic
matters.
\end{abstract}


\maketitle

\section{Introduction} Recent cosmic observations, including Type Ia Supernovae
\cite{ Riess,  Perlmutter1999}, Large Scale Structure  (LSS) \cite{Tegmark2004,
Tegmark2004a, Tegmark2006, Seljak2005,  AdelmanMcCarthy2006, Abazajian2005} and
Cosmic  Microwave Background  (CMB)  \cite{Spergel2007, Page2007,  Hinshaw2007,
Jarosik2007} have independently indicated that the evolution of the universe is
currently dominated by  a homogeneously distributed cosmic  fluid with negative
pressure, the so-called dark energy. The  dark energy fills the universe making
up of order $74\%$  of its energy budget. The remaining  energy fraction in the
present universe is about $22\%$ occupied  by the pressureless cold dark matter
and  $4\%$  occupied by  baryons.  Despite  many  years  of research  and  much
progress, the nature and the origin of dark energy does still remain as an open
issue.

Phenomenologically, the best candidate of dark energy that  fits perfectly the
observation  data  is  the  so-called  cosmological  constant  (CC)  $\Lambda$,
introduced  at first  by Einstein  in his  gravitational field  equations as  a
Lagrange  multiplier ensuring  the constancy  of the  4-volume of  the universe
\cite{Kunz2006,  Nobbenhuis2006}.   However,  CC  explanation  of  dark  energy
encounters  some   fundamental  obstacles   in  physics,  in   particular,  the
\emph{fine-tuning}     problem    and     the    \emph{coincidence}     problem
\cite{Amendola2010}.   From  the particle  physics  perspective,  CC should  be
interpreted as the  density of vacuum energy.  The physical  CC should contain
quantum corrections  from the  zero-point energies of  matter fields,  which is
close to  Planck density $M_{P}^{4}$  ($M_{P}= 1/\sqrt{8\pi G}$ is  the reduced
Planck mass)  in magnitude.  What  the fine-tuning problem  has to face  is the
great discrepancy between the observed value  of CC and its quantum correction,
the   former  is   only  $10^{-123}$   times   in  magnitude   as  the   latter
\cite{Kachru2000,   HenryTye2001,    Yokoyama2002,   Kane2003,   Mukohyama2004,
  Dolgov2008}.  Even  if   this  fine-tuning  problem  could   be  evaded,  the
coincidence problem as to why both energy densities of the observed dark energy
and the dark matter are of the same  order at present epoch remains, due to the
fact that CC is time independent and non-dynamical.

The resolution of fine-tuning problem has probably  to wait for the advent of a
satisfied quantum theory of gravity  in 4-dimensional spacetime. As a temporary
expedient in cosmology community, the physical  CC is assumed to vanish exactly
\cite{Deiss2012}   \footnote{This  assumption   does   actually  prohibit   the
interpretation  of CC  as the  energy  density of  vacuum fields.  CC does  not
fluctuate,  but  the vacuum  fields  fluctuate  and their  fluctuations  couple
universally to gravitation. We thank B. Deiss for pointing it to us.} and it is
conjectured that the dark energy which  drives the late time cosmic accelerated
expansion  is  dynamical.  The  observations  actually  say  little  about  the
evolution of  the equation  of state  (EoS) of  dark energy.   Researchers have
proposed   lots  of   alternative  models   (see  Reviews   \cite{Copeland2006,
Amendola2010, Cai2010a} and references therein) to explain the late time cosmic
acceleration  and   alleviate  the  corresponding  coincidence   problem.   The
well-known Chaplygin gas model  \cite{Kamenshchik2001} and its generalizations,
e.g., the models  proposed in Refs. \cite{Bento2002,  Bilic2002, Bilic2009} are
among these approaches.

The  Chaplygin gas  and its  generalizations  are of  the so-called  barotropic
fluids whose pressure depends only upon the energy density \cite{Amendola2010}.
What  distinguishes them  from other  dynamic dark  energy models  is that,  at
background level, they get rid of  the coincidence problem by unifying the dark
energy and  dark matter  into a  single dark  substance. The  gas behaves  as a
pressureless matter  for very large redshift,  however, it becomes a  CC as the
redshift  is  small.   Due  to  the  non-vanishing  squared   speed  of  sound,
unfortunately, the Chaplygin gas and its generalizations would have fatal flaws
for   explaining  tiny   but   stable   CMB  anisotropies   \cite{Amendola2003,
Sandvik2004}.  It follows  from Sandvik's  analysis in  Ref. \cite{Sandvik2004}
that  the unique  reliable Chaplygin  gas like  model is  the perfect  fluid of
constant pressure.   That the pressure of  a fluid is kept  invariant makes the
squared sound speed  vanish identically, resulting in the  disappearance of the
embarrassed oscillation-instability issue in density perturbations. The density
contrasts  of  constant-pressure  barotropic fluid  decay  monotonously  during
evolution, implying that the fluid does not unify the dark energy and cold dark
matter  into a  single dark  substance.  It  is merely  a mixture  of CC  and a
non-relativistic matter, and is equivalent  to the standard $\Lambda$-CDM model
to all  orders in perturbation  theory.  Consequently, the  coincidence problem
remains.

The fluid  models provide at  most some  effective descriptions of  the dynamic
dark energy.  In view of  the superstring/M-theory, a developing  but promising
candidate of quantum  gravity, the realization of dark energy  with some scalar
fields is  a more fundamental  description than  the perfect fluid  models. The
scalar fields are  ubiquitous in superstring/M-theory, they  arise naturally as
dilaton or the compactification moduli. A scalar field with a canonical kinetic
energy  is dubbed  \emph{quintessence} \cite{Peebles1988},  which is  among the
most studied  candidates for  dynamic dark  energy. The  quintessence typically
involves a  single scalar field  with a particular  self-interaction potential,
allowing the vacuum  energy to become dominant only  recently. The quintessence
potential has generally to be fine-tuned \emph{ad hoc} to solve the coincidence
problem.  At background level, a  quintessence field is generally equivalent to
a  perfect fluid  whose EoS  evolves  from \emph{plus  one} at  early times  to
approximately  \emph{minus   one}  at  present  epoch,   with  some  unphysical
ingredients introduced into the description of the evolution of Universe.

In this paper, we intend to study the cosmology of the quintessence counterpart
of a  perfect fluid  of constant  pressure. Different  from Chaplygin  gas like
models, a quintessence field can not be viewed as a barotropic fluid in general
\cite{Christopherson2009}. Whether the quintessence  field of constant pressure
is equivalent to $\Lambda$-CDM model or  provides a unified description to dark
energy and dark matter are open issues. Our main motivation in this paper is to
establish a  dynamic dark energy model  with a quintessence scalar  field whose
EoS  evolves from  zero at  early  times to  nearly  minus one  at present  and
potential  is not  required to  be fine-tuned  severely, so  that the  model is
brought back to life in search of solutions of the coincidence problem. We also
try to  examine the possibility to  build a unified description  of dark energy
and dark  matter in such  a scenario.  It is  shown that the  constant pressure
requirement enables the quintessence EoS interpolates between two plateaus, one
corresponds to cold dark  matter in the large redshift epoch,  another to CC as
the redshift comes near zero.  The quintessence potential comes completely from
the constant  pressure assumption.  It  is factorized  into the product  of the
squared Hubble  rate and  a dimensionless  quantity, dubbed  the \emph{reduced}
quintessence potential, which has also two  plateaus so that the fine-tuning of
the potential  is unnecessary.   Our investigation  in the  linear perturbation
theory  shows  that the  squared  sound  speeds of  quintessence  perturbations
oscillate around  \emph{unity}, and  the corresponding density  contrasts decay
monotonously  at late  times  during  their evolution.   We  conclude that  the
quintessence field of  constant pressure could only play the  role of a dynamic
dark energy,  which is neither a  unified dark matter  nor a mixture of  CC and
some non-relativistic matter. 

Throughout the paper we adopt the Planck units $c= \hbar = M_{P}=1$.

\section{Model} We begin  with the assumption that at the  background level our
universe is described by a flat Robertson-Walker metric
\begin{equation}
  \label{eq:1}
  ds^{2} = -dt^{2} + a^{2}(t) d\vec{x}^{2}
\end{equation} in which fills a mixture  of a quintessence scalar $\phi(t)$ and
a perfect fluid. The quintessence field is described by Lagrangian
\begin{equation}
  \label{eq:2}
  {\mathscr L} = -\frac{1}{2}\partial_{\mu}\phi \partial^{\mu}\phi -V(\phi)
\end{equation} but  the extra fluid  by its  pressure $P_f$ and  energy density
$\rho_f$.  There  are several  well-known candidates for  such a  cosmic fluid,
\emph{e.g.},  radiation,  baryonic  dust,  the  cold  dark  matter  (CDM),  the
cosmological  constant  $\Lambda$,  or  mixture  of them.   As  usual,  we  use
$\omega_f$  to denote  the EoS  of  this extra  fluid, defined  by $\omega_f  =
P_f/\rho_f$.   The quintessence  EoS  is similarly  defined  as $\omega_\phi  =
P_\phi/\rho_\phi$,   where    $P_\phi=\frac{1}{2}\dot{\phi}^2   -V(\phi)$   and
$\rho_\phi =  \frac{1}{2}\dot{\phi}^2 +V(\phi)$  are respectively  the pressure
and energy  density when  the quintessence  field is also  viewed as  a perfect
fluid. $\omega_\phi  \approx 1$  if kinetic  term dominates  while $\omega_\phi
\approx -1$ if  potential term dominates.  If both kinetic  and potential terms
are  almost  equivalently  important,  $\omega_\phi  \approx  0$.  It  is  also
remarkable  for  the  EoS  of  a  quintessence field  not  to  cross  over  the
cosmological  constant  boundary $\omega_{CC}=-1$  \cite{Vikman2005,  Kunz2006,
Cai2010a}.

In  the  flat  Robertson-Walker   background,  Einstein's  gravitational  field
equations become the so-called Friedmann equations:
\begin{eqnarray}
\label{eq:3} & & 3H^{2} = \frac{1}{2}\dot{\phi}^{2} + V(\phi) + \rho_f \\
\label{eq:4}  &  &  \dot{H}  =  -  \frac{1}{2}\dot{\phi}^{2}  -\frac{1}{2}  (1+
\omega_f) \rho_f
\end{eqnarray}  where  $H=\dot{a}/a$ is  the  Hubble's  expansion rate  of  the
universe for which we denote its present  value as $H_{0}$.  The extra fluid is
assumed not  to interact  with the quintessence  scalar field,  $\dot{\rho}_f +
3H(1+\omega_f) \rho_f =0$.  Consequently, the  evolution of the scalar field is
subject to the following Klein-Gordon equation,
\begin{equation}
\label{eq:5} \ddot{\phi} + 3H \dot{\phi} + V_{,\phi} = 0
\end{equation} which  is very  the energy conservation  equation of  the scalar
field at  the background  level. To  be convenient, we  will use  the so-called
\emph{efolding} number $N=\ln a$ as the  time variable from now on, $-\infty <N
<+\infty$, and let $N=0$ represent the  present epoch. A prime will represent a
derivative  with   respect  to   $N$,  unless  otherwise   specified.   Because
$t\rightarrow N=\ln a$ is merely a \emph{time transformation}, both coordinates
$(t, \vec{x})$ and $(N, \vec{x})$  are comoving ones \cite{Ellis2012}. In terms
of $N$, Eqs.(\ref{eq:3}), (\ref{eq:4}) and (\ref{eq:5}) are recast as:
\begin{eqnarray}
\label{eq:6} & &  3H^{2} = V(\phi) + \frac{1}{2}H^{2} \phi^{\prime  2} + \rho_f
\\
\label{eq:7} & & (H^2)^\prime = -H^2 \phi^{\prime 2} - (1+\omega_f) \,\rho_f
\end{eqnarray} and
\begin{equation}
\label{eq:8}     H^{2}    \phi^{\prime\prime}     +     \left[    V(\phi)     +
\frac{1}{2}(1-\omega_f)\rho_f \right] \phi^\prime + V_{,\phi}(\phi) = 0
\end{equation} respectively.

We want  to study  a quintessence  model in this  paper whose  EoS interpolates
between the pressureless matter dominant epoch  at large red shifts ($z \gtrsim
1$)\footnote{The red-shift parameter $z$ is defined as $z=e^{-N}-1$. Therefore,
$z\gtrsim 1$ corresponds to $N \lesssim  -0.693$.} and the dark energy dominant
epoch at  $z\approx 0$.  The  merits of such a  model are as  follows. Firstly,
there are few unphysical ingredients of EoS different from $\omega =1/3, 0$ and
$\omega  =-1$ involved  in  the alleviation  of the  coincidence  problem by  a
quintessence field.  Besides,  the model provides the possibility  to unify the
dark  energy and  dark  matter  into a  single  dark  substance, the  so-called
\emph{unified dark  matter} or \emph{quartessence} \cite{Makler2003}.   Here we
introduce  a  dimensionless quantity  $U(\phi)$,  referred  to as  the  reduced
quintessence potential from now on, so that $V(\phi)$ is factorized,
\begin{equation}
  \label{eq:9} V(\phi) = 3H^2 U(\phi)
\end{equation} Then we can translate Klein-Gordon equation (\ref{eq:8}) into:
\begin{equation}
  \label{eq:10}      H^2     \bigg[\frac{1}{2}\frac{d}{dN}      +     3U      +
\frac{3}{2}(1-\omega_f)  \Omega_f   \bigg]  \big[   \phi^{\prime  2}   -  6(1-U
-\Omega_f) \big] = 0
\end{equation} where $\Omega_f =\rho_f/{3H^2}$ is the reduced energy density of
the extra fluid. The first Friedmann equation (\ref{eq:6}) can equivalently be
expressed as,
\begin{equation}
  \label{eq:11} \frac{1}{2} \phi^{\prime 2} = 3(1-U-\Omega_f)
\end{equation}   In   terms    of   Eq.(\ref{eq:11}),   Klein-Gordon   equation
Eq.(\ref{eq:10}) turns out  to be an identity.  This is indeed the  case. As we
have explained,  among the  two Friedmann equations  (\ref{eq:6}), (\ref{eq:7})
and the Klein-Gordon equation (\ref{eq:8}), only two of them are independent.

What  looks  fascinating  here  is  that Eq.(\ref{eq:11})  is  similar  to  the
characteristic  equation of  an instanton  in  $(1+1)$ Euclidean  space if  the
\emph{potential}  $U(\phi)$  has  more   than  one  vacua  \cite{Rajaraman1982,
Vachaspati2006}. We  want to  know if the  cosmological observations  could put
some restrictions on the possible forms  of $U(\phi)$. With $U(\phi)$, the
energy density and pressure of the quintessence field are rewritten as,
\begin{eqnarray}
\label{eq:12}  & &  P_\phi  = H^{2}\left(  \frac{1}{2} \phi^{\prime  2}  - 3  U
\right)\\
\label{eq:13} &  & \rho_\phi =  H^{2}\left( \frac{1}{2}  \phi^{\prime 2} +  3 U
\right)
\end{eqnarray}
We can further simplify Eqs.(\ref{eq:12}) and (\ref{eq:13}) into:
\begin{eqnarray}
  \label{eq:14} & & \rho_\phi = 3H^2 (1-\Omega_f)\\
  \label{eq:15} & & P_\phi = 3H^2 (1-2U-\Omega_f)
\end{eqnarray}
by exploiting Eq.(\ref{eq:11}).  The cosmology of the model at the background
level is determined completely by quintessence  EoS $\omega_\phi$  and the
effective EoS parameter $\omega_{eff} = (P_\phi+P_f)/(\rho_\phi +\rho_f
) $ of the mixture fluid.  Obviously,
\begin{equation}
  \label{eq:16} \omega_\phi = 1- \frac{2U}{1-\Omega_f}
\end{equation}
and
\begin{equation}
  \label{eq:17} \omega_{eff}=1-2U-(1-\omega_f)\Omega_f
\end{equation} Provided  $\omega_{eff} \vert_{N\approx  0} <  -1/3$, the
late-time  accelerated  expansion  of  our  universe  occurs.   Therefore,  the
magnitude of the reduced quintessence potential  at the present epoch must obey
the    inequality:   $\frac{2}{3}    -   \frac{1}{2}(1-\omega_f)\Omega_f^{(0)}<
U\vert_{N\approx  0}$.   For  example,  if   the  extra  cosmic  fluid  is  the
non-relativistic matter, $\omega_f\approx  0$, $\Omega_f^{(0)}\approx 0.26$, we
have $U\vert_{N\approx 0} \gtrsim 0.54$.

The interpolation of  $\omega_\phi$ between zero and minus one  can be achieved
if the reduced quintessence potential $U(\phi)$  has at least two plateaus, one
is at  $U\approx \frac{1}{2}[1-(1-\omega_f)\Omega_f]$  for the  large red-shift
epoch and  another is at $U\approx  1 - \frac{1}{2} (1-  \omega_f) \Omega_f$ as
the red-shift parameter $z$ approaches to  zero.  To this end, the quintessence
field under consideration is defined to  have a vanishing adiabatic sound speed
$c^2_{\phi,  ~{ad}} =  {\partial P_\phi}/{\partial  \rho_\phi}$.  In  other
words, the pressure $P_\phi$ is supposed  not to change during the evolution of
the universe. Keeping $P_\phi$ invariant implies that,
\begin{equation}
  \label{eq:18} \omega_\phi^\prime = 3 \omega_\phi (1+ \omega_\phi)
\end{equation}  Obviously, the  expected  plateaus in  the $\omega_\phi\sim  N$
curve are guaranteed by this condition,  one is $\omega_\phi =0$ and another is
$\omega_\phi  = -1$.  Eq.(\ref{eq:18}) can  be translated  into the  definition
equation for the reduced quintessence potential,
\begin{equation}
  \label{eq:19} U^\prime + 3 \big[(1-U)(1-2U) - (1-U+ \omega_f U)\Omega_f \big]
= 0
\end{equation} The  quintessence field  is expected  to play  the role  of dark
energy  at low  red shifts  to excite  the late  time cosmic  acceleration. The
magnitude  of  the reduced  potential  $U$  must  be larger  than  $\frac{1}{2}
(1-\Omega_f)$ so that $P_\phi <0$. For  this scalar behaves as the pressureless
non-relativistic matter at large red shifts, the Hubble rate $H$ have to ascend
sharply with respect to the increase of the red-shift parameter.

In the absence of the extra fluid, $\Omega_f=0$, Eq.(\ref{eq:19}) can be solved
analytically, with a closed form solution given below:
\begin{equation}
  \label{eq:20}  U(N) =  \frac{1-\alpha  + (2\alpha  -1)e^{3N}}{2  -2 \alpha  +
(2\alpha -1)e^{3N}}
\end{equation} where $\alpha := U\vert_{N=0}$ is an integration parameter which
stands  for  the value  of  the  reduced  quintessence potential  $U(\phi)$  at
present. As pointed out previously, $\alpha  > 2/3$. Because $\alpha \ne 0$, we
see that: $1/2 \leq U(N) \leq 1$.

Substitution of Eq.(\ref{eq:20}) into  Eqs.(\ref{eq:11}) and (\ref{eq:7}) leads
to,
\begin{equation}
  \label{eq:21}           \phi(N)          =           \frac{2}{\sqrt{3}}\left[
\tanh^{-1}\frac{1}{\sqrt{2(1-\alpha)}}  - \tanh^{-1}  \sqrt{\frac{2(1-\alpha) +
(2\alpha-1)e^{3N} }{2(1-\alpha)}} \,\right]
\end{equation} and
\begin{equation}
\label{eq:22} H^2 = H_0^2\left[ (2\alpha -1) + 2(1-\alpha) e^{-3N} \right]
\end{equation} To guarantee the positivity  of $H^2$ during its evolution, $1/2
\leq  \alpha \leq  1$\footnote{When $\alpha=1$,  the quintessence  scalar field
under  consideration  degenerates  to  the cosmological  constant  $\Lambda  =3
H_0^2$.}. Therefore, the parameter $\alpha$ takes  its value in the region $2/3
<\alpha  \leq  1$.  Combination  of Eqs.(\ref{eq:21})  and  (\ref{eq:22})  with
Eq.(\ref{eq:9}) enables us to express the quintessence potential as an explicit
function of the scalar field itself. The result is,
\begin{equation}
  \label{eq:23} V(\phi)  = \frac{3H_0^2}{4} \left[ 3(2\alpha  -1) +(3- 2\alpha)
\cosh(\sqrt{3}\phi) -2 \sqrt{2(1-\alpha)} \sinh(\sqrt{3}\phi)\right]
\end{equation}

There  exist potentially  three different  interpretations of  the quintessence
field under consideration. The first possibility is simply to interpret it as a
dynamic dark  energy. Alternatively it  could be interpreted as  a unified dark
matter (quartessence). With these two interpretations, the quintessence field
is characterized by its energy density and pressure given below,
\begin{eqnarray}
\label{eq:24}  & &  \rho_\phi =  3H_0^2\left[(2\alpha -1)+  2(1-\alpha) e^{-3N}
\right]\\
\label{eq:25} & & P_\phi = -3H_0^2 (2\alpha -1)
\end{eqnarray} Though the pressure $P_\phi$  is a negative constant, the energy
density  $\rho_\phi$ interpolates  between that  of a  non-relativistic matter,
$\rho_\phi \approx  6H_0^2 (1-\alpha)  e^{-3N}$, for  $N \ll 1$  and that  of a
cosmological  constant, $\rho_\phi  \approx 3H_0^2  (2\alpha -1)$,  for $N  \gg
0$. The evolution of quintessence EoS is found to be,
\begin{equation}
  \label{eq:26}  \omega_\phi  =  -  \frac{(2\alpha  -1)  e^{3N}}{2(1-\alpha)  +
(2\alpha -1)e^{3N}}
\end{equation} In Figure 1, we plot  the evolution of $\omega_\phi$ by choosing
$\alpha =0.87$. As  expected, $\omega_\phi$ interpolates between  minus one for
$N \gtrsim 1$ and zero for $N \lesssim -2$.

\begin{figure}[!ht]
\label{wphi1}
\begin{center}
\includegraphics[angle=0]{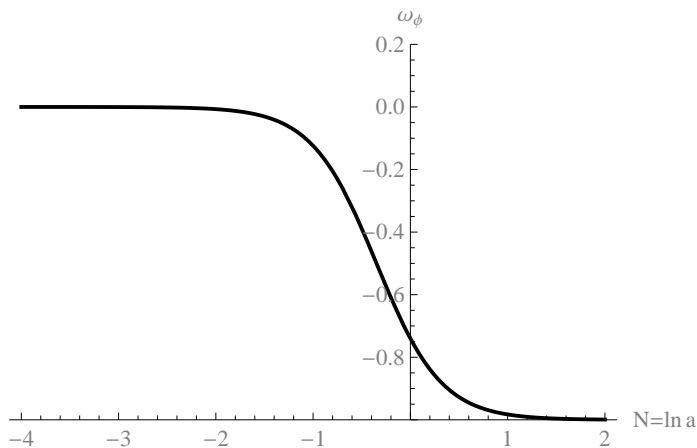} 
\caption[\emph{Evolution  of  energy  density against  $N$}]{Evolution  of  EoS
$\omega_\phi$ given  in Eqs.(\ref{eq:26})  for parameters  $\alpha= 0.87  $. As
expected, two  plateaus exist.  One plateau corresponds  to $\omega_\phi\approx
-1$ for $N  \gtrsim 1$, another plateau corresponds to  $\omega_\phi \approx 0$
for  $N \lesssim  -2$.  $\omega_\phi$ crosses  over  the acceleration  boundary
$-1/3$ at $N \approx -0.37$.}
\end{center}
\end{figure}

The  third   possibility  is   to  interpret   the  quintessence   field  under
consideration as a mixture of CC and some non-relativistic matter, $\rho_\phi =
\rho_{CC} +  \rho_{M}$. In this  scenario, the dark  energy is understood  as a
cosmological constant,  $\rho_{CC} = - P_{CC}  = 3H_0^2 (2\alpha -1)$,  but the
non-relativistic  matter  is characterized  by  $\rho_{M}  = 6H_0^2  (1-\alpha)
e^{-3N}$ and $P_{M} =0$.  The EoS parameters of two components are $\omega_{CC}
= -1$ and $\omega_{M} = 0$, respectively.  Because of $\omega_{CC} = -1$, there
is no interaction  between the components. The  quintessence EoS $\omega_\phi$,
as given in Eq.(\ref{eq:26}), should be understood as the effective EoS of such
a  mixture,  $\omega_\phi  =   \omega_{CC}\Omega_{CC}  +  \omega_M  \Omega_M  =
-\Omega_{CC}$, where the  dimensionless energy densities of  the components are
as follows,
\begin{eqnarray}
\label{eq:27}   &  &   \Omega_{CC}(N)  =\frac{(2\alpha   -1)}{(2\alpha  -1)   +
2(1-\alpha) e^{-3N}} \\
\label{eq:28}  & &  \Omega_{M}(N)  =\frac{2(1-\alpha)  e^{-3N}}{(2\alpha -1)  +
2(1-\alpha) e^{-3N}}
\end{eqnarray}  Notice  that  $\Omega_M(0)   =  2(1-\alpha)$.  Relying  on  the
observational constraint $\Omega_M(0) \approx 0.26$,  the best-fit value of the
parameter $\alpha$  might be  $\alpha \approx  0.87$. With  respect to  such an
interpretation,  the  quintessence  model   under  consideration  is  merely  a
reformulation of the standard $\Lambda$-CDM model.

All of three  interpretations of the quintessence field of  a constant pressure
appear  to  be  plausible  at  the background  level.  They  can,  however,  be
distinguished  from  each  other  if  we study  the  evolution  of  the  matter
perturbations. This will be addressed in the next section.

\begin{figure}[!ht]
\label{alphavalue}
\begin{center}
\includegraphics[angle=0]{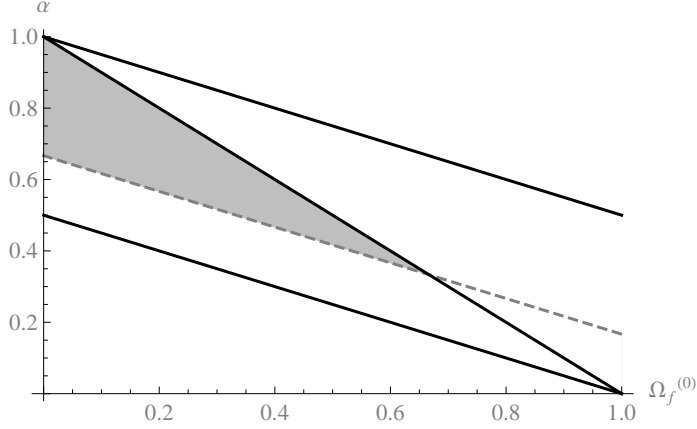} 
\caption[\emph{Dependence of $\alpha$  on $\Omega_f^{(0)}$}]{Parameter space of
$\alpha$ and $\Omega_{f}^{\left(0\right)}$. The solid black lines stand for the
critical  cases,   from  bottom   to  the  top   corresponding  to   $\alpha  =
(1-\Omega_{f}^{(0)})/2$,  $\alpha=1-  \Omega_{f}^{(0)}$  and   $\alpha  =  1  -
\Omega_{f}^{(0)}/2$   respectively.  The   dashed  line   corresponds  to   the
acceleration boundary $6\alpha -4 +  3\Omega_f^{(0)} =0$. All of the parameters
$(\Omega_f^{(0)}, \alpha)$ within  the region of gray shadow  above this dashed
line allow the existence of late time cosmic acceleration.}
\end{center}
\end{figure}

When the contribution of the extra cosmic fluid is taken into account,
\begin{equation}
\label{eq:29} \Omega_f = \frac{H_0^2}{H^2}\Omega_f^{(0)} e^{-3N (1+\omega_f)}
\end{equation} Eqs. (\ref{eq:7}) and (\ref{eq:11}) become,
\begin{eqnarray}
\label{eq:30} & & (H^2)^\prime + 6H^2(1-U) - 3H_0^2 \Omega_f^{(0)} (1-\omega_f)
e^{-3N(1+\omega_f)} = 0 \\
\label{eq:31} &  & \phi^{\prime  2} =  6(1-U) -6\Omega_f^{(0)}\frac{H_0^2}{H^2}
e^{-3N (1+\omega_f)}
\end{eqnarray}  respectively, and  the condition  (\ref{eq:19}) for  preserving
$P_\phi$ to be a constant during the evolution of the Universe is recast as:
\begin{equation}
  \label{eq:32}    U^\prime   +    3    \bigg[(1-U)(1-2U)   -    \Omega_f^{(0)}
\frac{H_0^2}{H^2}(1-U +\omega_f U) e^{-3N(1+\omega_f)} \bigg] = 0
\end{equation}  The closed-form  solutions to  Eqs.(\ref{eq:30}), (\ref{eq:31})
and (\ref{eq:32}) are not available in  general. However, if the extra fluid is
of  extremely non-relativistic,  $\omega_f  =0$, these  equations  can also  be
solved exactly. The corresponding solution is as follows:
\begin{eqnarray}
\label{eq:33}        &       &        \phi(N)       =        \frac{2}{\sqrt{3}}
\sqrt{\frac{2(1-\alpha-\Omega_f^{(0)})}{2-2\alpha-\Omega_f^{(0)}}}       \Bigg[
\tanh^{-1}\frac{1}{\sqrt{2-2\alpha-\Omega_f^{(0)}}}    \nonumber    \\   &    &
\,\,\,\,\,\,\,\,\,\,\,\,\,\,\,\,\,\,\,\,\,\,\,\,\,\,\,\,\,     -     \tanh^{-1}
\sqrt{\frac{2-2\alpha-\Omega_f^{(0)}   +  (2\alpha-1+\Omega_f^{(0)}   )  e^{3N}
}{2-2\alpha-\Omega_f^{(0)}}} \, \Bigg ]
\end{eqnarray}
\begin{equation}
\label{eq:34}  H^2 =  H_0^2 \left[(2\alpha  -1 +  \Omega_f^{(0)}) +  (2-2\alpha
-\Omega_f^{(0)}) e^{-3N}\right]
\end{equation}
\begin{equation}
\label{eq:35}    U    =    \frac{1-\alpha    -\Omega_f^{(0)}    +(2\alpha    -1
+\Omega_f^{(0)})e^{3N}}{2-2\alpha       -\Omega_f^{(0)}      +(2\alpha       -1
+\Omega_f^{(0)})e^{3N}}
\end{equation} where as before,  $\alpha =U\vert_{N=0}$.  Eq.(\ref{eq:35}) that
gives the closed form of the  reduced quintessence potential in the presence of
extra   fluid  does   differ  from   Eq.(\ref{eq:20})  but   recovers  it   for
$\Omega_f^{(0)} =0$.  Substitution of  Eqs.(\ref{eq:35}) and (\ref{eq:29}) into
(\ref{eq:16})  and (\ref{eq:17})  leads  to the  following  expressions of  the
quintessence EoS,
\begin{equation}
\label{eq:36}     \omega_\phi      =     -\frac{(2\alpha-1     +\Omega_f^{(0)})
e^{3N}}{2(1-\alpha -\Omega_f^{(0)})+(2\alpha-1 +\Omega_f^{(0)}) e^{3N}}
\end{equation} and the effective EoS of the mixture,
\begin{equation}
\label{eq:37}     \omega_{eff}     =     -\frac{(2\alpha-1     +\Omega_f^{(0)})
e^{3N}}{2-2\alpha -\Omega_f^{(0)}+(2\alpha-1 +\Omega_f^{(0)}) e^{3N}}
\end{equation}     Manifestly,     $\omega_\phi     \lesssim     \omega_{eff}$.
Eqs.(\ref{eq:34}), (\ref{eq:36}) and (\ref{eq:37}) implies that the behavior of
the  mixed  quintessence  field  and  the extra  fluid  is  determined  by  two
parameters,  i.e., the  initial  value of  the  reduced quintessence  potential
$\alpha$ and the initial dimensionless density $\Omega_{f}^{\left(0\right)}$ of
the  extra fluid.   The late  time accelerated  expansion occurs  if these  two
parameters  are  subject  to  the  inequality  $6\alpha  -4  +  3\Omega_f^{(0)}
>0$. Except for  $\alpha =1 -\Omega_f^{(0)}/2$, these two  parameters have also
to satisfy  the constraint  $\alpha \leq 1-  \Omega_f^{(0)}$ to  guarantee both
$\omega_\phi$  and  $\omega_{eff}$  nonsingular.  In  principle,  $\alpha$  and
$\Omega_f^{(0)}$ are two free parameters, however, there exist three degenerate
but important cases  where they are dependent upon one  another. The first case
corresponds  to   $\alpha=(1-\Omega_{f}^{(0)})/2$,  for   which  $\omega_{\phi}
=\omega_{eff}=0$, both  of the quintessence  field and  the extra fluid  act as
non-relativistic matters.   In the second  case, $\alpha=1-\Omega_{f}^{(0)}/2$,
$\omega_{eff}=-1$, the  mixture of the  quintessence field and the  extra fluid
behaves  as a  CC.  The  third degenerate  case is  characterized by  equations
$\alpha=1- \Omega_{f}^{(0)}$ and $\omega_{\phi}=-1$,  in which the quintessence
field  itself  is nothing  but  a  CC.  It  is  conspicuous  that the  standard
$\Lambda$-CDM is recovered in  the third case if we take the  values of the two
parameters as $\alpha \approx 0.74$ and $\Omega_f^{(0)} \approx 0.26$.

\begin{figure}[!ht]
\label{wphi2}
\begin{center}
\includegraphics[angle=0]{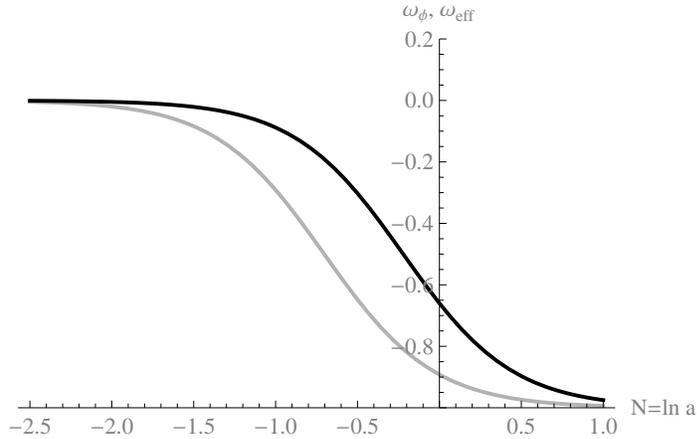} 
\caption{Evolution of  EoS $\omega_\phi$ and  $\omega_{eff}$ versus $N=  \ln a$
for   parameters  $\alpha=   0.7$,   $\omega_f=0$  and   $\Omega_f^{(0)}\approx
0.26$.  The  black curve  stands  for  $\omega_{eff}  $  while the  gray  curve
describes $\omega_\phi$.  Both curves interpolate  between two plateaus of zero
and of minus one.} 
\end{center}
\end{figure}

Because $\omega_\phi  \leq \omega_{eff}$,  it is  earlier for  $\omega_\phi$ to
cross the acceleration boundary of $\omega = -1/3$. In particular, if $\alpha =
1-  \Omega_f^{(0)}$,   $\omega_\phi  =   -1$  identically   but  $\omega_{eff}$
approaches to the onset of acceleration only at $N = - \frac{1}{3} \ln \big[2(1
-\Omega_f^{(0)})/\Omega_f^{(0)}\big]$.  Even if the quintessence field provides
a unified description for both dark energy and dark matter, it plays merely the
role of  dark energy  once $\omega_{\phi}$  crosses the  acceleration boundary.
Consequently,  the budget  of cold  dark  matter at  present epoch  have to  be
balanced by  some other cosmic  ingredients.  In this  paper, we use  simply an
extra fluid  of $\omega_f  =0$ to mimic  the mixture of  baryonic dust  and (at
least partial)  cold dark matter  so that  the observational constraint  on the
abundance   of  non-relativistic   matter  at   present  epoch   is  saturated,
$\Omega_f^{(0)} \approx 0.26$. For comparison we will consider four situations,
corresponding to $\alpha =0.62, ~0.66, ~0.70$ and $0.74$, respectively. As have
pointed out, the last case is a paraphrase of the standard $\Lambda$-CDM
model where the quintessence field behaves exactly as a cosmological constant,
$\omega_\phi =-1$.  The first three cases put forward models of dynamic dark
energy in  which the EoS of  the quintessence scalar field  interpolate between
two  plateaus of  zero and  of  minus one,  so that  the notorious  coincidence
problem with which the prototypic $\Lambda$-CDM model is embarrassed is greatly
alleviated. The existence of plateaus in  the $\omega\sim N$ curve implies that
the severe fine-tuning of the quintessence potential is not required.

\begin{table}[h]
\label{Table 1}
\centering
\begin{tabular}{p{55pt}p{40pt}p{40pt}p{40pt}p{40pt}p{40pt}}
\toprule
& $~~\alpha$ & $\Omega_f^{(0)}$ & $~\omega_{\phi}^{(0)}$  & $\omega_{eff}^{(0)}$ & $~~N_{c}$ \\
\midrule
Case 1 &  0.62 & 0.26 & -0.68 & -0.50 & -0.23\\
Case 2 &  0.66 & 0.26 & -0.78 & -0.58 & -0.34\\
Case 3 &  0.70 & 0.26 & -0.89 & -0.66 & -0.45\\
$\Lambda$-CDM &  0.74 & 0.26 & -1.00 & -0.74 & -0.58\\
\bottomrule
\end{tabular}
\caption{Parameter choice  for the  quintessence model under  consideration. We
use  $\omega_\phi^{(0)}$  and  $\omega_{eff}^{(0)}$  to denote  the  values  of
quintessence field  EoS and the  effective EoS  of the mixture  of quintessence
field and the  extra non-relativistic fluid at present epoch.  $N_c$ stands for
the onset for $\omega_{eff}$ crosses the acceleration boundary $-1/3$.}
\end{table}

\section{perturbations} Dark energy influences not  only the expansion rate, it
influences also  the growth rate of  matter perturbations. To clarify  what the
role does the quintessence field of constant pressure play during the evolution
of  the universe,  in this  section we  study the  matter perturbations  at the
linear  perturbation  level.   We  start with  the  perturbed  Robertson-Walker
metric,
\begin{equation}
  \label{eq:38}  ds^2 =  -(1+2\Psi) H^{-2}  dN^2 +  e^{2N}(1+2\Phi) \delta_{ij}
dx^i dx^j
\end{equation} in  Newtonian gauge  and for  convenience let  us work  out from
beginning a  single Fourier mode $k$  so that the perturbed  quantities $\Phi =
\Phi(N) e^{i\vec{k}  \cdot\vec{x}}$, $\Psi =  \Psi(N) e^{i\vec{k}\cdot\vec{x}}$
and  so  on.  The  perturbed  energy-momentum  tensor  of  the extra  fluid  is
described by the  perturbed energy density $\delta \rho_f  = - T^0_{0(f)}$,
the  perturbed pressure  $\delta P_f  = \frac{1}{3}\delta  T^i_{i(f)}$ and  the
velocity divergence  $\theta_f$ defined  by $i  k_j \delta  T^j_{0(f)} =  -(1 +
\omega_f)\rho_f  \theta_f$.   For  the  quintessence  field,   these  perturbed
quantities read:
\begin{eqnarray}
\label{eq:39} &  & \delta  \rho_\phi = H^2  (\phi^\prime \varphi^\prime  - \Psi
\phi^{\prime 2}) + V_{, \phi} \varphi \\
\label{eq:40}  & &  \delta  P_\phi  = H^2  (\phi^\prime  \varphi^\prime -  \Psi
\phi^{\prime 2}) - V_{, \phi} \varphi \\
\label{eq:41} & & \theta_\phi = \hat{\lambda}^{-2} \frac{\varphi}{\phi^\prime}
\end{eqnarray} where $\hat{\lambda}  = He^{N}/k$, and $\varphi  := \delta \phi$
denotes the field  fluctuation. The perturbation equations for  a perfect fluid
with density contrast $\delta_X$ and velocity divergence $\theta_X$ are:
\begin{eqnarray}
\label{eq:42} & & \delta^\prime_X = 3(\omega_X -c_{X}^2) \delta_X -(1+\omega_X)
(\theta_X + 3 \Phi^\prime) \\
\label{eq:43}  & &  \theta_X^\prime  =-  \frac{\theta_X}{2}\left[ 1-  6\omega_X
-3\omega_{eff}    +   \frac{2\omega_X^\prime}{1+\omega_X}    \right]   +
\hat{\lambda}^{-2} \left[ \frac{c_{X}^2 \delta_X}{1+ \omega_X} + \Psi \right]
\end{eqnarray}  where  $X$  is  either   $\phi$  or  $f$,  and  $c_X^2:={\delta
P_X}/{\delta \rho_X}$  the squared sound speed  of fluid $X$. The  evolution of
Bardeen potentials  $\Phi$ and  $\Psi$ is  subject to  Einstein's gravitational
equations. The result is,
\begin{eqnarray}
  \label{eq:44}
&  & \Phi = 3\hat{\lambda}^2 \left[ \frac{1}{2}\Omega_f \delta_f + \frac{1}{2}
  (1- \Omega_f) \delta_\phi + \Psi - \Phi^\prime \right] \\
\label{eq:45}
&  & \Phi^\prime = \Psi - \frac{3}{2}\hat{\lambda}^2 \bigg[ (1+ \omega_f)
\Omega_f \theta_f + (1+ \omega_\phi) (1- \Omega_f) \theta_\phi \bigg]  \\
\label{eq:46}
&  & \Psi = -\Phi
\end{eqnarray}  We  assume that  the  extra  fluid  is  of barotropic  so  that
$\omega_f = \omega_f(\rho_f)$ and $c_{f}^2 = dP_f/{d \rho_f}$. The quintessence
field, on the other hand, can not be regarded as a barotropic fluid in general.
When $X$  in Eqs.(\ref{eq:42})  and (\ref{eq:43})  stands for  the quintessence
field $\phi$ of constant pressure,  $\omega_\phi$ is given in Eq.(\ref{eq:16}),
but the squared  sound speed $c_{\phi}^2$ does not coincide  with its adiabatic
counterpart $c_{\phi, {ad}}^2$, 
\begin{equation}
  \label{eq:47}
  c_{\phi}^2 = \frac{H^2 (\phi^\prime \varphi^\prime -\Psi \phi^{\prime 2}) -
    V_{,\phi} \varphi}{H^2 (\phi^\prime \varphi^\prime -\Psi \phi^{\prime 2}) +
    V_{,\phi} \varphi}
\end{equation} In other words, $c_{\phi}^2$ depends on the detailed behavior of
both  perturbation   and  background  quantities.   There  is  no   reason  for
$c_{\phi}^2$  vanishes  identically.  In  fact,  if we  put  ourselves  in  the
quintessence  rest  frames,   we  have  $\varphi  =0$   and  hence  $c_{\phi}^2
=1$.  Without  loss   of  generality,  we  assume  $c_\phi^2  \ne   0$  in  the
following. Eq.(\ref{eq:43}) for $\theta_\phi$ turns  out to become an identity,
but Eq.(\ref{eq:42}) for $\delta_\phi$ reduces to:
\begin{equation}
  \label{eq:48}
  \varphi^{\prime\prime} + \left( 3 + \frac{H^\prime}{H} \right)
  \varphi^\prime + \left( \hat{\lambda}^{-2}
  + \frac{V_{,\phi\phi}}{H^2} \right) \varphi
  + 4 \phi^\prime \Phi^\prime  - 2 \frac{V_{,\phi}}{H^2} \Phi = 0
\end{equation}
where,
\begin{eqnarray}
\label{eq:49}
&   & V_{,\phi} = -\frac{3}{2}H^2 \phi^\prime \\
\label{eq:50}
&   & V_{,\phi\phi} =\frac{9}{4} H^2 (3-2U -\Omega_f)
\end{eqnarray} At late times ($z\lesssim 10$\footnote{i.e., $N \gtrsim -2.4$.})
all modes of interest have entered the comoving horizon \cite{Dodelson2003}. In
view of  observations, the typical scales  relevant to the galaxy  matter power
spectrum  correspond  to  the  wavenumbers   $k  \lesssim  0.1  h  ~{Mpc}^{-1}$
\cite{Sandvik2004}   or  equivalently   $k  \lesssim   300H_0$.   Consequently,
Eqs.(\ref{eq:42}),  (\ref{eq:43})   for  the  extra  fluid   perturbations  and
Eq.(\ref{eq:48}) for the quintessence fluctuations should be solved for $ 30H_0
\lesssim k  \lesssim 100 H_0$,  for which  $\hat{\lambda} \ll 1$.   Notice that
$\Phi \sim  {\mathscr O}(\hat{\lambda}^2)$  and $U  \sim {\mathscr  O}(1)$. The
coefficient of field fluctuation $\varphi$  in Eq.(\ref{eq:48}) is dominated by
$\hat{\lambda}^{-2}$  when  $\hat{\lambda}  \ll  1$,   which  is  the  case  on
sub-horizon scales ($k \gg H_0 $) if $N$ is not very negative.  Let us consider
temporarily the  case of $\Omega_f  =0$ and solve  numerically Eq.(\ref{eq:48})
with initial conditions  $\varphi\vert_{N=0}=0$ and $\varphi^\prime \vert_{N=0}
=1$. The  solution tells that the  field fluctuations oscillate around  zero on
sub-horizon   scales,  with   nearly   vanishing  average   values  over   many
oscillations. As illustrated in Figure 4, the field fluctuations of sub-horizon
modes ($30H_0 \lesssim  k \lesssim 300H_0$) could approximately  be regarded as
zero  for $N  \gtrsim -3$.  This conclusion  can even  be extrapolated  to some
earlier time,  say $N \gtrsim -5$,  with slightly poor accuracy.  It holds also
for parameters $\alpha$ and $\Omega_f^{(0)}$  assuming other values as in Table
1.   Consequently,   the  velocity  divergence  $\theta_\phi$   of  sub-horizon
quintessence modes oscillates around zero and its sound speed oscillates around
one, and  for $N  \gtrsim -5$,  we have $\theta_\phi  \approx 0$  and $c_\phi^2
\approx 1$.   The approximate equality  $\theta_\phi \approx 0$ for  $N \gtrsim
-5$ further  implies the  decoupling of Eq.(\ref{eq:42})  from Eq.(\ref{eq:43})
for  $X=\phi$.  The  former  turns  out to  become  a first-order  differential
equation of $\delta_\phi$ for $\hat{\lambda} \ll 1 $,
\begin{equation}
\label{eq:51} \delta_\phi^{\prime} + 3(1- \omega_\phi) \delta_\phi -\frac{9
  H^2}{2k^2} e^{2N}(1+ \omega_{\phi})\delta_{\phi} = 0
\end{equation} The solution of Eq.(\ref{eq:51}) takes the following closed form
as $k \rightarrow \infty$,
\begin{equation}
\label{eq:52}   \delta_\phi  =   \frac{\delta_\phi(0)  e^{-3N}}{2(1-\alpha)   +
(2\alpha -1) e^{3N} }
\end{equation} Unfortunately, such a density  contrast falls sharply before the
universe enters the  late-time accelerated expansion. It  contributes little to
the galaxy clustering.   This is the same as the  sub-horizon density contrasts
of  the barotropic  fluid of  constant pressure  \cite{Sandvik2004}.  In  fact,
$\delta_\phi$ vanishes identically for $\alpha=0.74$ and $\Omega_f^{(0)} =0.26$
because  in this  case the  quintessence  field plays  the  role of  a pure  CC
\footnote{When  $\omega_\phi   =-1$,  Eq.(\ref{eq:51})  is  solved   by  either
$\delta_\phi  \sim e^{-6N}$  or $\delta_{\phi}  =0$. However,  the former  is a
specious solution because it conflicts with Eq.(\ref{eq:43}) for $X=\phi$ under
the assumptions  of $\theta_\phi  \approx 0$ and  $c_\phi^2 \approx  1$.}.  The
quintessence  field of  constant pressure  could not  take charge  of structure
formation even if it unifies the dark energy and some (fuzzy) dark matter \footnote{According to Eq.(\ref{eq:50}), the mass of quintessence field reads,
\[
m_\phi = \sqrt{V_{,\phi\phi}} =\frac{3}{2}H_0\sqrt{2\alpha +\Omega_f^{(0)} -1 + 2(2-2\alpha -\Omega_f^{(0)}) e^{-3N}}
\]
For the parameters given in Table 1, say $\alpha =0.62$ and $\Omega_f^{(0)} = 0.26$, we have $m_\phi\vert_{N\approx -5} \approx 2700 H_0 \approx 4\times 10^{-30}$ eV, which is much less than the possible mass limit ($m_\nu < 1.43$ eV) of the massive neutrinos \cite{Kristiansen2007}.} into a single
dark substance.   Notice that  the sound speed  of the mixture  of CC  and some
non-relativistic  matter vanishes  identically  \cite{Amendola2010}.  That  the
physical sound speed of quintessence field does not vanish obviously mismatches
this  fact,   which  expels  definitely   the  possibility  to   interpret  the
quintessence field of constant pressure as a mixture of CC and non-relativistic
matters. It is better to interpret  the quintessence field of constant pressure
as a dynamic dark energy, it unifies at  most the dark energy and a fraction of
(fuzzy)  dark matter.   The quintessence  field clusters,  but such  a clustering
decays monotonously during the evolution that it takes no direct responsibility
for the  structure formation on large  scales.  The structure formation  in the
proposed scenario must have other impellers.

\begin{figure}[!ht]
\label{fieldfluctuation}
\begin{center}
\includegraphics[angle=0]{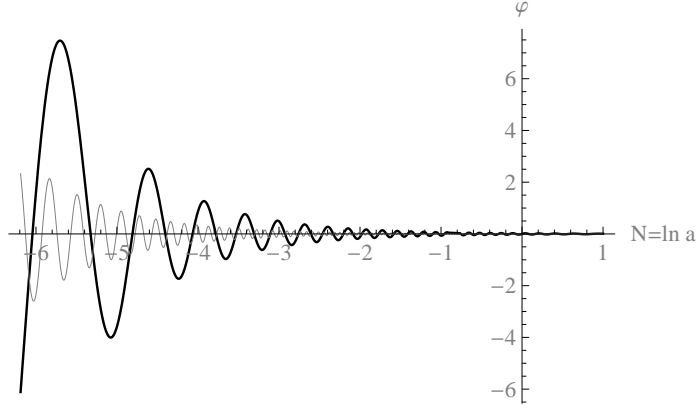} 
\caption[\emph{Evolution of  field fluctuation against $N$}]  {Evolution curves
of field  fluctuations $\varphi$ of sub-horizon  modes in the absence  of extra
fluid.  The curves are obtained from the numerical solution of Eq.(\ref{eq:48})
for $k=40H_0$ (black) and $k=160H_0$ (gray).  The other parameters are taken to
be $\alpha  =0.87$ and $\Omega_f^{(0)}  =0$.  The field fluctuations  evolve as
damped oscillations around  zero, and their amplitudes  vanish effectively when
$N \gtrsim -3$ (or $N \gtrsim -5$ with slightly poor accuracy). }
\end{center}
\end{figure}

The cold  dark matter which is  believed to answer for  the structure formation
can effectively  be described as  an extremely non-relativistic  perfect fluid.
We should  take $\Omega_f \ne  0$ and $\omega_f =0$  in our model  building. In
such  a  scenario, the  extra  fluid  is a  mixture  of  cold dark  matter  and
baryons.  Though Eq.(\ref{eq:48})  describes still  damped oscillations  around
zero so that  $\varphi \approx 0$ for  $N$ not to be very  negative ($N \gtrsim
-5$) and  the fluctuations of  quintessence field decay on  sub-horizon scales,
the sub-horizon fluctuations  of the non-relativistic fluid  grow steadily.  It
follows from Eqs.(\ref{eq:42}), (\ref{eq:43}), (\ref{eq:44}), (\ref{eq:45}) and
(\ref{eq:46}) that, for $\hat{\lambda} \ll 1$,
\begin{eqnarray}
\label{eq:53}
&   & \delta_\phi^{\prime} + a_0 \delta_\phi
= b_0 \delta_f + b_1 \delta_f^\prime  \\
\label{eq:54}
&   & \delta_f^{\prime\prime} + c_1\delta_f^\prime + c_0 \delta_f = d_0
\delta_\phi + d_1 \delta_\phi^\prime
\end{eqnarray}
where,
\begin{eqnarray}
\label{eq:55}
&   & a_0 = 3(1- \omega_\phi) + \frac{9}{2}\hat{\lambda}^2(1+
\omega_\phi)(1-\Omega_f) \\
\label{eq:56}
&   & b_0 = \frac{9}{2}\hat{\lambda}^2 (1+ \omega_\phi) \Omega_f \\
\label{eq:57}
&   & b_1 = -\frac{9}{2}\hat{\lambda}^2 (1+ \omega_\phi) \Omega_f \\
\label{eq:58}
&   & c_0 = -\frac{3}{2}\Omega_f -\frac{9}{4}\hat{\lambda}^2 \bigg[2\Omega_f^\prime
-3\Omega_f^2 -(1+ 9 \omega_{eff}) \Omega_f \bigg] \\
\label{eq:59}
&   & c_1 = \frac{1}{2}(1-3\omega_{eff}) + \frac{9}{2}\hat{\lambda}^2
\bigg[\Omega_f^\prime -(1+ 3\omega_{eff}) \Omega_f \bigg ] \\
\label{eq:60}
&   & d_0 = \frac{3}{2}(1-\Omega_f) -\frac{9}{4}\hat{\lambda}^2
\bigg[2\Omega_f^\prime + 3\Omega_f (1-\Omega_f) + (1+9 \omega_{eff}) (1-
\Omega_f) \bigg]
\\
\label{eq:61}
&   & d_1 = \frac{9}{2}\hat{\lambda}^2 (1-\Omega_f)
\end{eqnarray}  In obtaining  these  coefficients  the assumption  $\theta_\phi
\approx 0$ on sub-horizon scales has been made use of, which could be justified
only when $N$ is not very negative.   If we take the extreme sub-horizon limit,
$k\rightarrow   \infty$,  we   can  further   simplify  Eqs.(\ref{eq:53})   and
(\ref{eq:54}) as,
\begin{eqnarray}
\label{eq:62} & & \delta_\phi^{\prime} + 3(1-\omega_\phi) \delta_\phi = 0 \\
\label{eq:63} & & \delta_f^{\prime\prime} + \frac{1}{2}(1-3\omega_{eff})
\delta_f^\prime   -\frac{3}{2}\Omega_f   \delta_f   =   \frac{3}{2}(1-\Omega_f)
\delta_\phi
\end{eqnarray}Eq.(\ref{eq:62}), as expected, describes a decaying field density
contrast $\delta_\phi$. Eq.(\ref{eq:63}),  on the other hand,  allows a growing
density  contrast  $\delta_f$  for   matter  perturbations.   It  follows  from
Eq.(\ref{eq:63})  that  the  quintessence  density contrast  plays  a  role  of
external   source  to   the  evolution   of   the  density   contrast  of   the
non-relativistic   fluid.   Provided   this   external   source   is   ignored,
Eq.(\ref{eq:63}) can approximately be solved by a closed-form solution,
\begin{eqnarray}
\label{eq:64}
& \delta_f & = C_1 e^{3\gamma_- N}~_2F_1(\gamma_{-}, \gamma_{+}+1/2,
1+\gamma_{+}+\gamma_{-}; \xi )  \nonumber \\
&   & ~~ + C_2 e^{-3\gamma_{+} N}~_2F_1 (- \gamma_{-}+1/2, -\gamma_{+},
1-\gamma_{+}-\gamma_{-}; \xi)
\end{eqnarray}
where $C_1$ and $C_2$ are two integration constants,
\begin{equation}
  \label{eq:65}
  \gamma_{\pm} = \frac{1}{12}\left[ \sqrt{\frac{2\alpha
      -23\Omega_f^{(0)}-2}{2\alpha +\Omega_f^{(0)} -2}} \pm 1 \right ],~~~~~
\xi = \frac{2\alpha +\Omega_f^{(0)} -1}{2\alpha  +\Omega_f^{(0)} -1}e^{3N}
\end{equation}   and   $_2F_1(a,b,c;\xi)$   stands   for   the   hypergeometric
function.  In case $\alpha =1-\Omega_f^{(0)}$,  the quintessence  field
degenerates  to  a  CC  (i.e., $\delta_\phi  =0$) and the  solution  given  in
Eq.(\ref{eq:64}) becomes of exact,
\begin{equation}
  \label{eq:66}
  \delta_m =C_1 e^N ~_2F_1
  \left(
  \frac{1}{3},1,\frac{11}{6};\frac{\alpha e^{3N}}{\alpha-1}
  \right) + C_2 e^{-3N/2} ~_2F_1
  \left(
  \frac{1}{6},-\frac{1}{2},\frac{1}{6};\frac{\alpha e^{3N}}{\alpha-1}
  \right)
\end{equation}

\begin{figure}[!ht]
\label{densitycontrast}
\begin{center}
\includegraphics[angle=0]{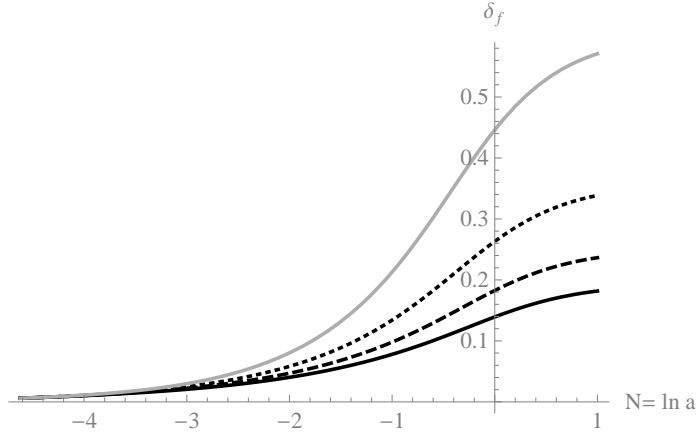} 
\caption[\emph{Evolution   of  density   contrast   of  matter   perturbation}]
{Evolution of density contrasts of  matter perturbations of sub-horizon mode $k
\approx 40 H_0$. The curves in  the bottom-up order describe the matter density
contrast  $\delta_f$  as  the   numerical  solution  of  Eqs.(\ref{eq:53})  and
(\ref{eq:54}) for  $\alpha = 0.62,  ~0.66, ~0.70$ and $0.74$,  respectively. In
all cases the  parameter $\Omega_f^{(0)}$ is assumed to be  $0.26$. Among these
curves the gray  one corresponds to the sub-horizon matter  density contrast in
the  standard  $\Lambda$-CDM model.  The  initial  conditions are  assigned  as
$\delta_f\vert_{N=-4.62} \approx  \delta_f^\prime\vert_{N=-4.62} \approx 0.006$
and  $\delta_\phi\vert_{N=-4.62} \approx  10^{-6}$.  The  larger the  parameter
$\alpha$  is, the  faster the  matter density  contrast grows  in the  proposed
scenario.  }
\end{center}
\end{figure}

If the  external source term is  taken into account, a  closed-form solution to
Eq.(\ref{eq:63}) seems to be impossible. Furthermore, the fact that the typical
smallest  scale relevant  to the  galaxy matter  power spectrum  corresponds to
$k\approx  300H_0$ requires  us  to solve  Eqs.(\ref{eq:53}) and  (\ref{eq:54})
instead for  sub-horizon modes satisfying observational  constraint $k \lesssim
300H_0$ \cite{Sandvik2004}. The  key scale in the matter power  spectrum is the
matter-radiation equality scale $k_{eq}\sim 100$  Mpc (i.e., $k_{eq} \approx 40
H_0$), which  defines the turn-around in  the spectrum. These equations  can be
solved numerically if some appropriate initial conditions are assigned.  We use
Eqs.(\ref{eq:53}) and (\ref{eq:54}) to  evolve the perturbations from $N\approx
-4.62$\footnote{i.e.,  $z\approx  100$.  This  is  a  typical redshift  in  the
cosmological epoch  dominated by  non-relativistic matter.}  until  today.  The
numerical     solution     of     $\delta_f$    with     initial     conditions
$\delta_f\vert_{N=-4.62} \approx  \delta_f^\prime\vert_{N=-4.62} \approx 0.006$
and $\delta_\phi\vert_{N=-4.62}  \approx 10^{-5}$  is plotted  in Figure  5 and
Figure 6  for the  typical mode  $k\approx 40H_0$, where  we take  the relevant
parameters $\alpha$ and $\Omega_f^{(0)}$ as  in Table 1.  The hierarchy between
the        initial        conditions       $\delta_f\vert_{N=-4.62}$        and
$\delta_{\phi}\vert_{N=-4.62}$  has  been   fine-tuned  so  that  $\delta_\phi$
vanishes  effectively   as  the  quintessence   field  mimics  only  a   CC  in
$\Lambda$-CDM model.  As expected,  in all cases  the matter  density contrasts
grow steadily while  the quintessence density contrasts decline  for $N \gtrsim
-3$. Except  for the $\Lambda$-CDM  model where $\delta_\phi$ falls  sharply to
zero, the quintessence density contrasts  undergo a period of increase preceded
their final-stage decline. This probably is a  clue that there could not be any
observable Sachs-Wolfe  effect \cite{Dodelson2003}  \cite{Amendola2010} related
to  the  quintessence  fluctuations  provided we  fine-tune  appropriately  the
initial condition of  $\delta_\phi$. From Figure 5 we see  that the sub-horizon
matter density contrasts in the proposed scenario of dynamic energy models grow
remarkably slower  than that in  $\Lambda$-CDM model.   Due to the  feedback of
non-vanishing  quintessence  density  contrasts,   see  Figure  6,  the  matter
perturbations in these models enter  the accelerated expansion epoch later than
those  in $\Lambda$-CDM.   The onset  of nonlinearity  occurs when  the density
perturbations obey  $\delta_f \gtrsim 0.3$.  For  a comoving scale of  100 Mpc,
i.e., $k\approx  40H_0$, this occurs  at redshift $z^*\sim  1$ \cite{Ellis2012}
which  corresponds to  $N^* \approx  -0.693$.  It  follows from  Figure 5  that
$\delta_f\vert_{N^*} \approx 0.09,  ~0.12, ~0.17$ in the  proposed scenarios of
$\alpha \approx0.62, ~0.66,  ~0.70 $ and $\delta_f\vert_{N^*}  \approx 0.28$ in
the standard $\Lambda$-CDM model, respectively.  The matter density contrast as
a function of comoving wavenumber at $N^*$ is also shown in Figure 7.

\begin{figure}[!ht]
\label{perturbation6}
\begin{center}
\includegraphics[angle=0]{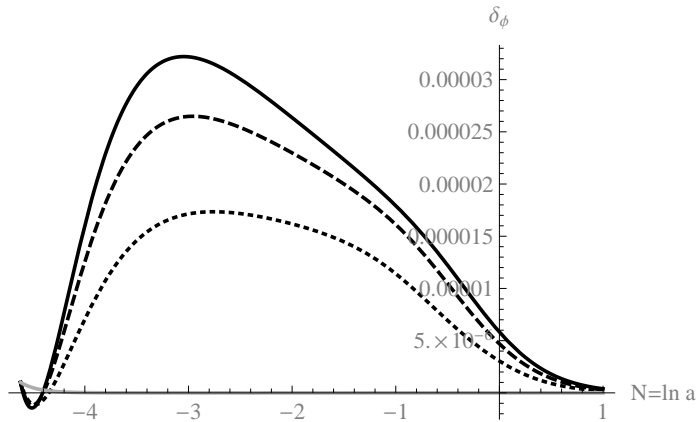} 
\caption[\emph{Evolution of  growth rates of matter  perturbations}] {Evolution
of quintessence density  contrasts of sub-horizon mode $k \approx  40 H_0$. The
parameter choice is  the same as Figure  5, so are the  initial conditions. The
curves correspond to $\alpha  = 0.62, ~0.66, ~0.70$ and $0.74$  from the top to
the  bottom. In  the first  cases the  quintessence density  contrasts increase
during the period $-4.4 \lesssim N \lesssim -3$ and then decrease gradually. In
the last case which corresponds  to $\Lambda$-CDM model, the quintessence field
is almost equivalent to a cosmological  constant and its density contrast drops
sharply to zero.  The decrease of  $\delta_\phi$ as $N \gtrsim -3$ excludes the
possibility for the considered quintessence field  to unify dark energy and the
cold dark matter into a single dark  substance. On the other hand, the increase
of $\delta_\phi$ prior  to $N \approx -3$  implies that there could  not be any
observable Sachs-Wolfe effect  related to the quintessence  fluctuations if the
initial condition for $\delta_\phi$ is fine-tuned appropriately.}
\end{center}
\end{figure}

\begin{figure}[!ht]
\label{perturbation7}
\begin{center}
\includegraphics[angle=0]{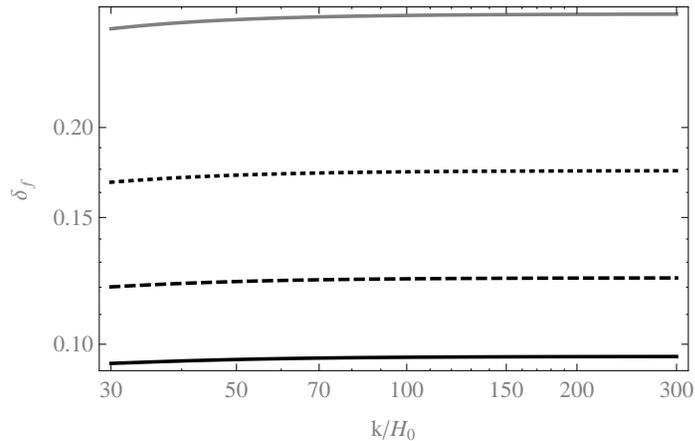} 
\caption[\emph{Evolution of growth rates  of matter perturbations}] {The matter
density  contrasts on  sub-horizon scales  at  the onset  of nonlinearity  $N^*
\approx -0.693$. The parameter choice is  as Table 1.  The curves correspond to
$\alpha = 0.62, ~0.66, ~0.70$ and $0.74$ from the bottom upward.  }
\end{center}
\end{figure}

The growth  of matter  perturbations can  be measured  by the  so-called growth
rate, which  is defined as  the ratio of the  derivative of the  matter density
contrast  with respect  to enfolding  time to  matter density  contrast itself,
i.e., $f = {\delta^\prime_f}/{\delta_f}$ in  the present scenario. In Figure 8,
we  plot the  evolution of  the growth  rates of  matter perturbations  of mode
$k\approx 40H_0$ for $\alpha =0.62, ~0.66$ and $0.70$ as well as their analogue
in $\Lambda$-CDM.  Because  the cosmological constant does not  cluster, it has
no  influence  on  the  evolution  of matter  perturbations,  in  the  standard
$\Lambda$-CDM  model the  growth  rate of  matter  perturbation remains  almost
invariant before it  enters acceleration. If the role of  dark energy is played
by the quintessence scalar field  of constant pressure, however, the clustering
of  quintessence  perturbations does  certainly  affect  the growth  of  matter
perturbations.   In the  proposed scenario,  although the  quintessence density
contrast does  not take  charge of the  large scale structures,  it acts  as an
external  driving  force in  the  evolution  equation (\ref{eq:54})  of  matter
density contrast.  Such an influence results in remarkable decline in $f\sim N$
curve prior to the epoch of acceleration. As shown in Figure 8, the smaller the
parameter $\alpha$ is, the greater  the quintessence field contributes to (fuzzy)
dark matter  and the faster  the sub-horizon perturbations  of non-relativistic
matters  decline.   When  the  cosmological  acceleration  begins,  $N  \gtrsim
-0.693$, the magnitude of $\delta_\phi$ drops effectively to zero and no longer
provides robust driving force, so the  growth rates of matter perturbations ebb
steadily in all cases.

\begin{figure}[!ht]
\label{growthrate}
\begin{center}
\includegraphics[angle=0]{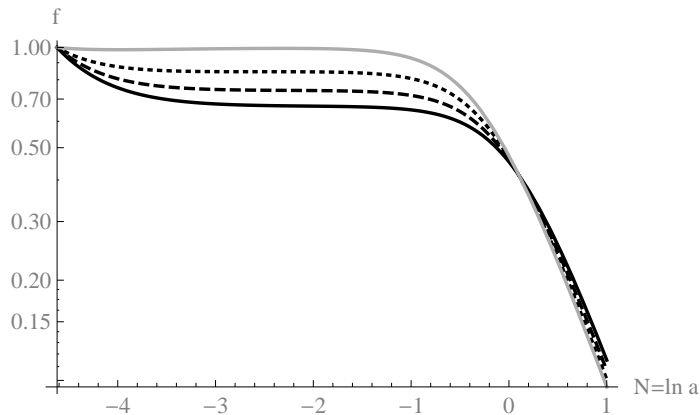} 
\caption[\emph{Evolution of  growth rates of matter  perturbations}] {Evolution
of growth rates  of matter perturbations in the  scenarios under consideration.
Black curves  describe the growth rates  of the density contrast  $\delta_f$ of
matter perturbations of $k\approx 40H_0$ for $\Omega_f^{(0)} =0.26$ and $\alpha
=0.62$ (solid), $0.66$ (dashed) and  $0.70$ (dotted), respectively.  Gray curve
on the  top is  the growth  rate of  the matter  density contrast  of $k\approx
40H_0$ in the standard $\Lambda$-CDM model. }
\end{center}
\end{figure}

\section{Conclusion}  In  this  paper  we  have  studied  the  cosmology of  a
quintessence scalar field which can be viewed as a non-barotropic perfect fluid
of constant pressure in linear perturbation theory.  The assumption of constant
pressure ensures the interpolation of  the quintessence EoS between the plateau
of zero  at large red shifts  and that of minus  one at small red  shifts.  Not
this characteristic alleviates  greatly the coincidence problem  only, it brings
also  few  unphysical   ingredients  into  the  description   of  evolution  of
universe. The potential of the quintessence field is determined completely from
the  requirement of  constant pressure,  which  is unnecessary  to be  severely
fine-tuned in the alleviation of coincidence problem. The adiabatic sound speed
of the quintessence  field vanishes, however, its physical  sound speed depends
upon the  details of perturbations  and equals  effectively to unity  for modes
within the  deep comoving horizon.  The  quintessence field does not  unify the
dark energy and the whole cold dark  matter into a single dark substance, it is
also not a mixture  of these two dark substances. What the  role it really play
during evolution is  plausibly a dynamic dark energy that  could cluster.  This
clustering  is measured  by  a decaying  density contrast.   Though  it is  not
directly responsible to the formation  of large scale structure, the clustering
of the quintessence perturbations could  effectively alleviate the growth rates
of matter perturbations as an external driving force.

\section*{Acknowledgments}
We would like to thank Y. F. Cai, M. Z. Li, J. X. Lu and
M. L. Yan for valuable discussions. The work was supported in part by NSFC under
No. 11235010.

\section*{References}
\bibliography{DarkEnergy}
\bibliographystyle{utcaps}
\end{document}